\documentclass[aps,prb,
 amsmath,amssymb,
 twocolumn, groupedaddress,
 showpacs]{revtex4-1}
\usepackage{bm}
\usepackage{graphicx}
\usepackage{gensymb}
\usepackage{amsmath}
\usepackage{amssymb}
\usepackage{color}
\usepackage{ulem}
\usepackage{here}

\begin{document}

\title{Evidence for bulk nodal loops and universality of Dirac-node arc surface states in ZrGeX$_{\rm c}$ (X$_{\rm c}$ = S, Se, Te)}
\author{Takechika Nakamura,$^1$ Seigo Souma,$^{2,3}$ Zhiwei Wang,$^4$ Kunihiko Yamauchi,$^5$ Daichi Takane,$^1$ Hikaru Oinuma,$^1$ Kosuke Nakayama,$^1$ Koji Horiba,$^6$ Hiroshi Kumigashira,$^{6,7}$ Tamio Oguchi,$^5$ Takashi Takahashi,$^{1,2,3}$ Yoichi Ando$^4$, and Takafumi Sato$^{1,2,3}$}

\affiliation{$^1$Department of Physics, Tohoku University, Sendai 980-8578, Japan\\
$^2$Center for Spintronics Research Network, Tohoku University, Sendai 980-8577, Japan\\
$^3$WPI Research Center, Advanced Institute for Materials Research, Tohoku University, Sendai 980-8577, Japan\\
$^4$Physics Institute II, University of Cologne, K\"oln 50937, Germany\\
$^5$Institute of Scientific and Industrial Research, Osaka University, Ibaraki, Osaka 567-0047, Japan\\
$^6$Institute of Materials Structure Science, High Energy Accelerator Research Organization (KEK), Tsukuba, Ibaraki 305-0801, Japan\\
$^7$Institute of Multidisciplinary Research for Advanced Materials (IMRAM), Tohoku University, Sendai 980-8577, Japan
}

\date{\today}

\begin{abstract}
We have performed angle-resolved photoemission spectroscopy (ARPES) on layered ternary compounds ZrGeX$_{\rm c}$ (X$_{\rm c}$ = S, Se, and Te) with square Ge lattices. ARPES measurements with bulk-sensitive soft-x-ray photons revealed a quasi-two-dimensional bulk-band structure with the bulk nodal loops protected by glide mirror symmetry of the crystal lattice. Moreover, high-resolution ARPES measurements near the Fermi level with vacuum-ultraviolet photons combined with first-principles band-structure calculations elucidated a Dirac-node-arc surface state traversing a tiny spin-orbit gap associated with the nodal loops. We found that this surface state commonly exists in ZrGeX$_{\rm c}$ despite the difference in the shape of nodal loops. The present results suggest that the spin-orbit coupling and the multiple nodal loops cooperatively play a key role in creating the exotic Dirac-node-arc surface states in this class of topological line-node semimetals.
\end{abstract}


\maketitle

\section{INTRODUCTION}
Topological insulators (TIs) are characterized by a gapless edge or a surface state (SS) protected by time-reversal symmetry [\onlinecite{HasanRMP2010, ZhangRMP2011, AndoJPSJ2013}]. The discovery of TIs triggered extensive explorations of new types of topological phases of matter, such as topological crystalline insulators with surface Dirac-cone bands protected by mirror reflection symmetry [\onlinecite{FuPRL2011, HsiehNC2012, TanakaNP2012, XuNC2012, DziawaNM2012}] or time-reversal-invariant topological superconductors with surface Majorana modes traversing the bulk superconducting gap [\onlinecite{LFuPRL2010, SasakiPRL2011, NMRNP2016, SasakiPRL2012, Sato2017}]. While all these topological materials commonly possess a finite energy gap in the bulk low-energy excitations, there exists another class of topological materials called topological semimetals (TSMs) which are characterized by gapless bulk excitations [\onlinecite{WengCM2016, FangCPB2016, WangAPX2017, ArmitageRMP2018, BernevigJPSJ2018, MurakamiJPSJ2018, SchoopCM2018, YangARX2018}]. Currently, TSMs are a target of intensive investigations, since they provide a promising platform to realize various exotic quantum phenomena distinct from those of topological materials with a bulk gap.

TSMs are classified into two categories depending on how the bulk valence and conduction bands contact each other in the Brillouin zone (BZ). Dirac semimetals (DSMs) and Weyl semimetals (WSMs) are characterized by linearly dispersing bulk bands crossing at a discrete single point in $k$ space (Dirac/Weyl point). On the other hand, in line-node semimetals (LNSMs), the crossing point extends one-dimensionally in $k$ space (nodal line/loop), being protected by the crystal symmetry such as mirror reflection or nonsymmorphic glide mirror symmetry. It has been theoretically predicted that WSMs host Fermi-arc SSs which connect the surface projection of Weyl points [\onlinecite{WanPRB2011, YangPRB2011, HaldaneArXiv2014, MurakamiPRB2014}]. In fact, the existence of such Fermi-arc SSs was verified by angle-resolved photoemission spectroscopy (ARPES) with noncentrosymmetric transition-metal monopnictides such as TaAs, NbAs, or NbP [\onlinecite{XuSci2015, BQLuPRX2015, YLChenNP2015, BQLuPRL2015, XuPRL2016, SoumaPRB2016}]. In contrast, no unified explanation has been established on how the SSs emerge in LNSMs because the symmetry associated with the bulk line node is often broken at the surface and, as a result, there is no guarantee for the emergence of robust topological SSs. Hence, it is an important challenge to identify and understand the SSs in LNSMs.

Among various LNSMs, particular attention has been paid to the layered ternary system {\it WHM} ({\it W} = Zr and Hf; {\it H} = Si, Ge, Sn, and Sb; {\it M} = O, S, Se, and Te) with the PbFCl-type (space group P4/$nmm$) crystal structure [see Fig. 1(a)]. Previous ARPES studies on ZrSiS with vacuum-ultraviolet (VUV) photons [\onlinecite{SchoopNC2016, NeupanePRB2016, ChenPRB2017}] identified two types of line nodes; one below the Fermi level ($E_{\rm F}$) extending along the {\it XR}/{\it MA} line [see Fig. 1(b)] protected by screw and time-reversal symmetries, and another with a diamond shape around $E_{\rm F}$ at $k_z$ = 0 and $\pi$ planes protected by glide mirror symmetry. Besides these bulk states, a SS has been commonly observed around the $\bar{X}$ point in ZrSiX$_{\rm c}$ (X$_{\rm c}$ = S, Se, Te), HfSiS, ZrSnTe, and ZrGeTe [\onlinecite{SchoopNC2016, NeupanePRB2016, LouPRB2016, TakanePRB2016, ToppNJP2016, HosenPRB2017, ChenPRB2017, floatingPRX2017, HosenPRB2018}]. This SS is located away from the nodal loops and may originate from a dangling bond or a ``floating'' SS [\onlinecite{floatingPRX2017}] created by a symmetry reduction at the surface, as shown in first-principles band-structure calculations [\onlinecite{ZrSiSCalc2015}]. Intriguingly, it was discovered that there exists another, unexpected SS alongside the nodal loops in HfSiS [\onlinecite{TakanePRB2016}]. This SS shows a peculiar {\it Dirac-node-arc} dispersion with the Dirac point extending one-dimensionally along the $\overline{\Gamma}\overline{M}$ line, and is neither reproduced by the bulk-band- nor the slab-calculations. Since its discovery in 2016, this Dirac-node-arc SS attracts a lot of attention [\onlinecite{floatingPRX2017, HosenPRB2018, Kumar2017}] as a rare case where a topological feature is discovered {\it without} any theoretical prediction. To pin down the origin of this unexpected SS, it is of great importance to clarify whether a similar SS exists or not in other {\it WHM} compounds and how its properties are linked to the bulk nodal features.

In this article, we report comprehensive ARPES studies of ZrGeX$_{\rm c}$ (X$_{\rm c}$ = S, Se, and Te). By utilizing bulk-sensitive soft-x-ray (SX) photons, we identify a quasi two-dimensional (2D) Fermi surface (FS) hosting bulk nodal loops at $k_z$ = 0 and $\pi$. The ARPES with VUV photons revealed a Dirac-node-arc SS dispersing across the small spin-orbit gaps of nodal loops at $k_z$ = 0 and $\pi$; importantly, this SS is commonly found in all three compounds, demonstrating its universal nature. We discuss characteristics of the observed SS in relation to the bulk nodal loops characterized by the Zak phase and the SSs found in other LNSMs.
 
 \section{EXPERIMENTAL AND THEORETICAL METHODS}
 
High-quality single crystals of  ZrGeX$_{\rm c}$ (X$_{\rm c}$ = S, Se, and Te) were synthesized with a chemical vapor transport method [\onlinecite{TakanePRB2016}]. ARPES measurements with VUV photons were performed with a Scienta-Omicron SES2002 electron analyzer with synchrotron light at BL28 in Photon Factory. We used circularly polarized light of 36-200 eV. ARPES measurements with bulk-sensitive SX photons were performed at BL2 with 350-600 eV photons with horizontal linear polarization. The energy resolutions for VUV- and SX-ARPES measurements were set at 10-30 meV and 150 meV, respectively. Samples were cleaved {\it in situ} along the (001) crystal plane in an ultrahigh vacuum of 1$\times$10$^{-10}$ Torr, and kept at 30 K during the measurements. Sample orientation was determined by Laue x-ray diffraction prior to the ARPES experiment. The Fermi level of samples was referenced to that of a gold film evaporated onto the sample holder. We have confirmed that the cleaved surface contains a shiny mirror-like flat plane, as can be recognized from a sharp laser spot reflected from the cleaved sample surface. The cleaved surface is always terminated by the X$_{\rm c}$ atoms, which is consistent with our observation that the spectral feature is fairly reproducible for different cleaves.

First-principles band-structure calculations were carried out by a projector augmented wave method implemented in Vienna Ab initio Simulation Package (VASP) code  [\onlinecite{VASP}] with generalized gradient approximation (GGA) [\onlinecite{GGA}]. The spin-orbit coupling (SOC) was included as the second variation in the self-consistent-field iterations. For the bulk calculations, in-plane lattice constant is fixed to the experimental value. For the slab calculations, we have chosen a four unit cell (four quintuple layers; QLs) of ZrGeX$_{\rm c}$ and adopted a periodic slab model with 1.5-nm-thick vacuum layer. Lattice parameter and atomic positions have been relaxed during the slab calculation except for the in-plane lattice constant which is fixed to the experimental value.

 \begin{figure}
\begin{center}
\includegraphics[width=3.4in]{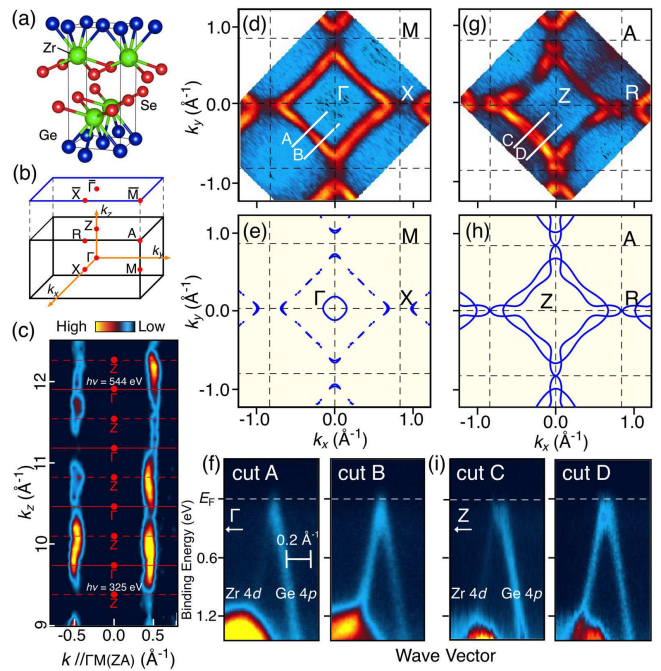}
  \hspace{0.2in}
\caption{(color online). (a) Crystal structure of ZrGeSe. (b) Bulk tetragonal BZ (black) and corresponding surface BZ (blue). (c) ARPES-intensity mapping at $E_{\rm F}$ of ZrGeSe as a function of $k_z$ and in-plane wave vector $k_{\parallel}$ along $\Gamma \it M$($\it ZA$) line measured at $h\nu$ = 300-600 eV. (d) ARPES-intensity mapping at $E_{\rm F}$ as a function of in-plane wave vectors ($k_x$ and $k_y$) at $k_z$ = 0 ($h\nu$ = 544 eV). (e) Calculated Fermi surface for ZrGeSe at $k_z$ = 0. (g) Same as (d) but at $k_z$ = $\pi$ ($h\nu$ = 325 eV). (h) Same as (e) but at $k_z$ = $\pi$. (f), (i) ARPES-intensity plots as a function of in-plane wave vector and $E_{\rm B}$ along cuts A and B in panel (d) and cuts C and D in panel (g), respectively.
}
\end{center}
\end{figure}

\begin{figure*}
\includegraphics[width=7 in]{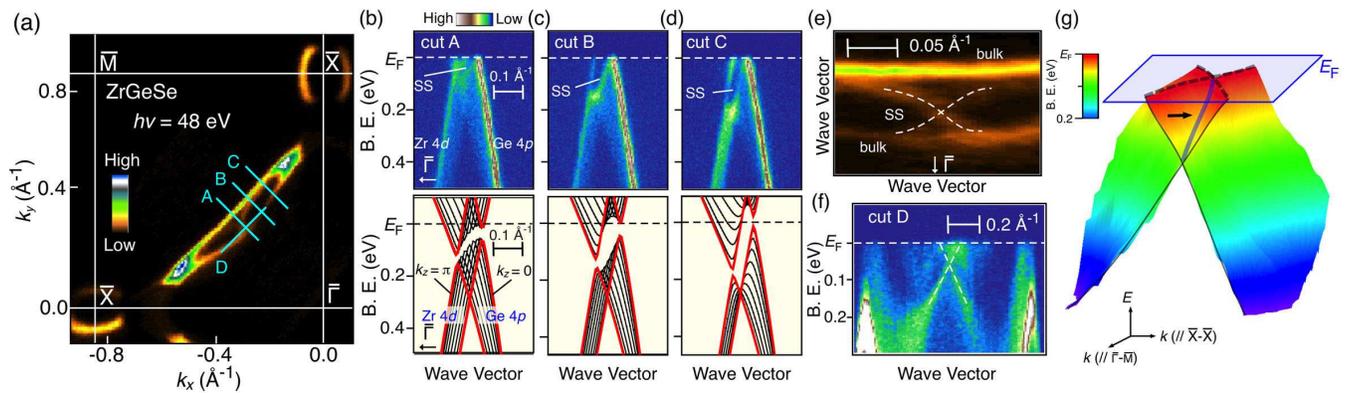}
  \hspace{0.2in}
\caption{(color online).(a) ARPES-intensity mapping at $E_{\rm F}$ at $h\nu$  = 48 eV. (b)-(d) ARPES intensity near $E_{\rm F}$ along cuts A-C (top) and corresponding calculated bulk-band dispersions for several $k_z$'s ranging from 0 to $\pi$ with the interval of 0.1$\pi$. (e) Magnified view of the banana-shaped FS measured at $h\nu$ = 48 eV. Dashed line is a guide for the eyes to trace the FS of SSs. (f) Near-$E_{\rm F}$ ARPES intensity along cut D in (a). (g) Schematic band structure in $E$-$k$ space for the Dirac-node-arc SS obtained from the experimental band dispersions around the banana-shaped FS.
}
\end{figure*}

 \section{RESULTS AND DISCUSSION}

 First, we present the bulk-band structure of ZrGeSe. To visualize the nodal loops originating from the bulk bands, it is essential to use bulk-sensitive SX photons. In fact, previous VUV-ARPES measurements on this material class ($e.g.$ ZrSiS and HfSiS) [\onlinecite{SchoopNC2016, NeupanePRB2016, LouPRB2016, TakanePRB2016, ToppNJP2016, HosenPRB2017, ChenPRB2017, floatingPRX2017, HosenPRB2018}] significantly suffered from the strong $k_z$-broadening effect and, as a result, could not precisely argue the three-dimensional (3D) dispersion of the bulk bands. Figure 1(c) shows the ARPES-intensity mapping at $E_{\rm F}$ as a function of $k_z$ and in-plane wave vector $k_{\parallel}$ along the $\overline{\Gamma}\overline{M}$ cut obtained with SX photons. One can recognize a nearly straight bright intensity pattern at $\sim\pm$0.5 \AA$^{-1}$ which originates from the FS formed by the Ge 4{\it p} and Zr 4{\it d} bands. This FS shows a finite wiggling with a periodicity matching that of the bulk BZ. These results establish the quasi-2D nature of the electronic states in ZrGeSe. 

To visualize the FS topology associated with the predicted nodal loop at $k_z$ = 0, we chose the photon energy of 544 eV and mapped out the ARPES intensity as a function of the in-plane wave vectors in the $\Gamma${\it XMX} ($k_z$ = 0) plane, as shown in Fig. 1(d) (for the relationship with the nodal loops and calculated band structure, see Fig. 6 in Appendix A). The obtained large diamond-like FS centered at $\Gamma$ shows a good agreement with the calculated FS, which includes SOC [Fig. 1(e)], reflecting the nodal loop near $E_{\rm F}$ protected by nonsymmorphic glide mirror symmetry of the crystal (note that the small circular pocket at $\Gamma$ predicted in the calculation is absent in the experiment). In fact, as shown in Fig. 1(f), the energy dispersion along the $k$ cut passing through the nodal loop [cut A; see also Fig. 1(d)] shows a $\Lambda$-shaped dispersion due to the intersection of the Zr 4{\it d} and Ge 4{\it p} bands. This means that the observed loop at $E_{\rm F}$ is essentially an extended version of the ``Dirac point'' coming from an extended band crossing. Such a feature is also resolved along the cut B, consistent with the existence of a nodal loop formed by the continuous band crossing in $k$ space. It is noted that the nodal loop is actually gapped with maximally 30 meV upon inclusion of SOC in the calculation, as detailed later. 
 
Figures 1(g) and 1(h) display the ARPES-intensity mapping for {\it ZRAR} ($k_z$ = $\pi$) plane measured at $h\nu$ = 325 eV and the corresponding calculated FS, respectively. One can see that the shape of experimental FS at $k_z$ = $\pi$ slightly shrinks compared to that at $k_z$  = 0, with a stronger concave distortion. Importantly, the FS at $k_z$ = $\pi$ seems to have a ``neck'' near {\it R}, which is absent in the FS at $k_z$ = 0. Such spectral feature is consistent with the calculated FS shown in Fig. 1(h), although the experiment does not resolve two diamond-like FS sheets centered at $Z$, mainly due to the lower resolution of the SX-ARPES compared to the VUV one. It is noted that this dual FS is caused by a downward energy shift of the nodal loop and the resultant $E_{\rm F}$ crossing of the upper Dirac-like band (for details, see Appendix A). Nevertheless, a $\Lambda$-shaped Dirac-like band can be visualized along the cuts C and D in Fig. 1(i), in support of the existence of a nodal loop on the $ZRAR$ plane. These results support the existence of nodal loops at both $k_z$ = 0 and $\pi$ as the sole energy states near $E_{\rm F}$ in ZrGeSe. We will discuss later that existence of two nodal loops is essential for the emergence of the Dirac-node-arc SS.

\begin{figure}
 \includegraphics[width=3.2in]{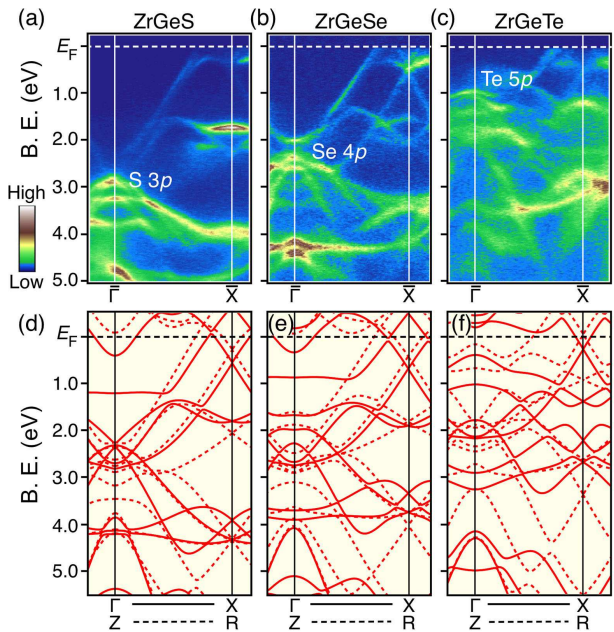}
 \hspace{0.2in}
\caption{(color online). (a)-(c) Comparison of ARPES-derived band dispersions among ZrGeX$_{\rm c}$ (X$_{\rm c}$ = S, Se, and Te), together with (d)-(f) corresponding calculated band structure at $k_z$ = 0 (solid curve) and $\pi$ (dashed curve).
}
\end{figure}

Now that the bulk nodal-loop structure is established, the next important issue is to clarify the existence of a SS. For this sake, we performed surface-sensitive VUV-ARPES measurements. Figure 2(a) displays the ARPES-intensity mapping at $E_{\rm F}$ as a function of in-plane wave vectors for ZrGeSe measured at $h\nu$ = 48 eV. One finds a ``banana''-shaped FS elongated along the $\overline{XX}$ direction, together with a small pocket at $\bar{X}$, similarly to ZrSiS and HfSiS [\onlinecite{SchoopNC2016, NeupanePRB2016, TakanePRB2016, ChenPRB2017, floatingPRX2017}]. The banana-shaped FS is associated with the bulk nodal loop while the small pockets at $\bar{X}$ originate from the surface floating band [\onlinecite{floatingPRX2017, FuSXarXiv2017}]. As seen in Fig. 2(b) where the ARPES intensity (top) along the cut A [marked in Fig. 2(a)] is compared with the corresponding bulk band-structure calculations (bottom), the energy bands that contribute to the banana-shaped intensity [Fig. 2(a)] originate mainly from the linearly dispersing Zr 4{\it d} band at $k_z$ = $\pi$ and the Ge 4{\it p} band at $k_z$ = 0. This assignment is consistent with the Fermi vectors in Figs. 1(f) and 1(i). One can also see in Fig. 2(b) a broad and weak intensity pattern inside these bands at the binding energy $E_{\rm B}$ of 0.1-0.5 eV, which shows a good correspondence with the projection of the calculated Zr 4{\it d} and Ge 4{\it p} bands. These results confirm that the spectral feature observed with VUV-ARPES significantly suffers a strong $k_z$-broadening effect [\onlinecite{SchoopNC2016, NeupanePRB2016, LouPRB2016, TakanePRB2016, ToppNJP2016, HosenPRB2017, ChenPRB2017, HosenPRB2018}].
 
 We found a signature of the SS inside the banana-shaped FS. As visible in the ARPES intensity along the cut B [Fig. 2(c)], there exists an unusual band dispersion within 0.15 eV of $E_{\rm F}$ whose band velocity is obviously different from that of the bulk bands. This band is likely attributed to the SS [\onlinecite{TakanePRB2016}], since it is not reproduced in the bulk-band calculations [bottom panel of Fig. 2(c)]. The absence of such feature in more bulk-sensitive SX-ARPES may be in support of its surface origin, although poorer resolution of SX-ARPES could also be responsible for it. In addition, the band seems to exist inside the projected gap of the bulk bands [see also Fig. 2(d)]. A close look at the banana-shaped FS in Fig. 2(e) further reveals that the SS forms a cross-shaped FS inside the banana [Fig. 2(e)] with an X-shaped dispersion along the cut D [Fig. 2(f)]. Such a behavior is better visualized in the experimental band dispersion in the $k_x$-$k_y$ space as shown in Fig. 2(g), which signifies the X-shaped dispersion extending along a line on the $\overline{\Gamma}\overline{M}$ plane forming a characteristic Dirac-node arc [\onlinecite{TakanePRB2016}].
 
 Next, to examine the universality of the spectral feature among the ZrGeX$_{\rm c}$ family, we have systematically performed VUV-ARPES measurements of ZrGeX$_{\rm c}$ (X$_{\rm c}$ = S, Se, and Te), and compare in Fig. 3 the experimental valence-band dispersion measured along the $\overline{\Gamma}\overline{X}$ line, together with the calculated band dispersions. One can notice several dispersive bands in all the compounds. Holelike bands topped at $\sim$3 eV in ZrGeS are attributed to the S 3$p$ bands. One can also recognize similar chalcogen bands in ZrGeSe [Fig. 3(b)] and ZrGeTe [Fig. 3(c)], while these bands systematically move upward from S- to Te-compounds. Such a chemical trend is qualitatively reproduced in the calculations shown in Figs. 3(d)-3(f), although some quantitative discrepancies are seen in the energy position of the chalcogen bands. As seen in Fig. 3(c), the Te 5{\it p}bands in ZrGeTe are relatively close to $E_{\rm F}$ in the experiment and they complicate the band structure near $E_{\rm F}$. Moreover, the ARPES intensity of ZrGeTe is relatively broad compared to those of ZrGeS and ZrGeSe. This could originate from the larger $k_z$ dispersion, as inferred from the large difference in the energy location of the calculated bands between $k_z$ = 0 and $\pi$ [Fig. 3(f)] compared to those in ZrGeS and ZrGeSe [Figs. 3(d) and 3(e)]. We also note that while the energy position of the chalcogen {\it p} bands changes significantly, the states near $E_{\rm F}$ are rather unchanged, as seen in the similar ARPES intensity profile between ZrGeS and ZrGeSe [Figs. 3(a) and 3(b)].   
 
With this clarification of the chalcogen-dependence of the valence-band structure, next we clarify the evolution of the FS. Figures 4(a)-4(d) show a direct comparison of the FS mapping among ZrGeX$_{\rm c}$ family and HfSiS [\onlinecite{TakanePRB2016}]. One can immediately notice that the FS topology is essentially the same among these compounds except for small quantitative differences. For example, a small pocket from the trivial SS at $\bar{X}$ is commonly resolved, whereas the doubling of this FS due to the Rashba splitting seen in HfSiS [\onlinecite{TakanePRB2016}] is not clearly observed in ZrGeX$_{\rm c}$, likely because of a weaker atomic SOC of Zr compared to that of Hf. The banana-shaped FS is also resolved in all the compounds, while that of ZrGeTe appears to be relatively ``fat'' and the corner of the banana near $\bar{X}$ is obscured. This difference may be attributed to the Te 5{\it p} bands, which are located closer to $E_{\rm F}$ and have a larger $k_z$ dispersion in ZrGeTe as seen in Fig. 3. Despite the quantitative difference in the shape of the banana-like FS and the bulk-band dispersion, Fig. 4(e) shows that the unusual band-bending near $E_{\rm F}$ commonly occurs in all the compounds, demonstrating that the Dirac-node-arc SS is a universal feature of ZrGeX$_{\rm c}$.
 
 \begin{figure*}
\includegraphics[width=5in]{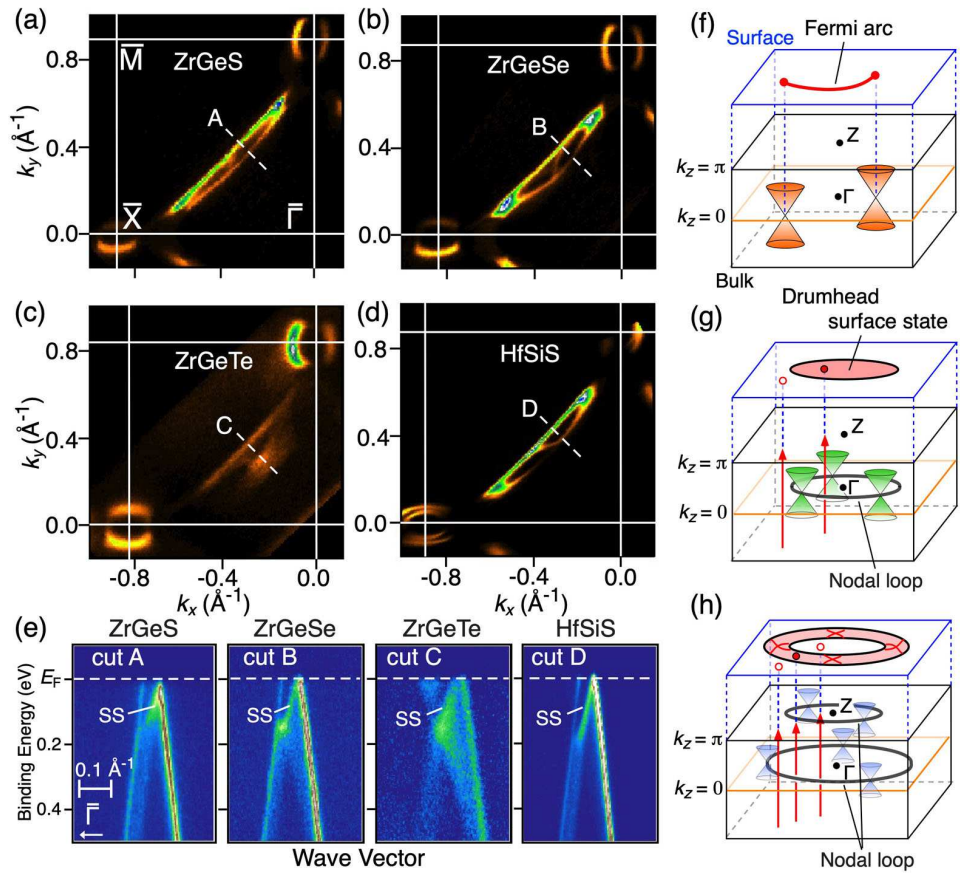}
\hspace{0.2in}
\caption{(color online). (a)-(d) ARPES-intensity mapping at $E_{\rm F}$ of (a) ZrGeS, (b) ZrGeSe, (c) ZrGeTe, and (d) HfSiS \cite{TakanePRB2016}. (e) Near-$E_{\rm F}$ ARPES intensity along cuts A-D in panels (a)-(d), respectively. (f)-(h), Comparison of schematic surface and bulk electronic structure among WSM, LNSM with single nodal loop and drumhead SS, and ZrGeX$_{\rm c}$. Red vertical allow along $k_z$ shows an integral path of Berry connection. Red filled and open circles on surface BZ stand for the Zak phase of $\pi$ and 0, respectively.
}
\end{figure*}

Now we discuss the possible origin of the Dirac-node-arc state. One may wonder that this feature might originate from the bulk band, since it lies at the edge of the bulk-band projection, as suggested from a comparison of the calculated band structure and the ARPES intensity in Figs. 2(b)-2(d). However, this is unlikely the case because (i) one can see a linearly dispersive bulk band up to $E_{\rm F}$ besides the band forming the Dirac-node arc, and (ii) the band degeneracy along $\overline{\Gamma}\overline{M}$ [see Fig. 2(g)] is not reproduced in the bulk-band calculations. The surface origin of the Dirac-node arc is in line with recent scanning-tunneling-microscopy experiments, which suggest the existence of SSs around the $\overline{\Gamma}\overline{M}$ line based on the quasiparticle interference pattern [\onlinecite{LodgeNL2017, SuNJP2018}]. Provided that the Dirac-node arc is a SS, one may naturally think that it would be reproduced in the slab calculations. However, while our slab calculations appear to show energy bands traversing the bulk-band gap which might be assigned as the surface states, these bands show ordinary dispersion in the $k_x$-$k_y$ plane, different from the dispersion expected from the Dirac-node arc (for details, see Appendix B). Although we carried out the slab calculations by taking into account various possibilities ({\it e.g.}, stacking faults, surface relaxations, topmost unit layers which are isolated from the bulk, surface islands with dispersive edge-localized states), none of them satisfactorily reproduced the observed Dirac-node-arc dispersion (for details, see Appendix C). Moreover, when SOC is included in the slab calculations, it always leads to an opening of a finite spin-orbit gap along whole $\overline{\Gamma}\overline{M}$ line, and therefore the symmetries peculiar to the surface do not allow a gapless band crossing along this line, which contradicts the experiment. One may argue that the electron correlations and the resultant band-mass renormalization of quasiparticles, revealed in the quantum-oscillation experiments [\onlinecite{ZrSiSNP2017, RudenkoPRL2018}], may be enhanced at the surface, but the generality of such a correlation-induced mass enhancement is questionable. We think that the reason why the Dirac-node-arc SS is hardly reproduced by the slab calculations is partially because of the very small bulk spin-orbit gap. This would lead to longer penetration depth of the SS into bulk, and could promote the strong hybridization between the top and bottom SSs to result in the modification of the shape of calculated SS; however such hybridization should not occur in a real crystal (for details, see Appendix C).

Nevertheless, it would be useful to summarize the similarities and differences of the observed SS with those of other LNSMs to deepen its understanding. It is well known that Fermi-arc SSs connecting the surface projection of Weyl nodes exist in WSMs due to the topological requirement, which essentially arises from the nature of Weyl-node pairs having opposite topological charges [Fig. 4(f)]. On the other hand, in LNSMs, there is no unified picture regarding the relationship between the nodal lines and the appearance of a SS because of the difficulty in explicitly defining the topological properties for various types of line nodes [\onlinecite{HirayamaNC2017, AhmPRL2018, FangPRB2015}]. In some LNSMs with a single nodal loop such as CaAgP [\onlinecite{YamakageJPSJ2016}] and Ca$_3$P$_2$ [\onlinecite{ChanPRB2016}], theories predicted the existence of drumhead SSs in the band-inverted $k$ region that connect the surface projection of nodal points between positive and negative $k$'s across the $\Gamma$ point [Fig. 4(g)] [\onlinecite{BurkovPRB2011, KimPRL2015, BianPRB2016}]. On the other hand, in ZrGeX$_{\rm c}$, the SS connects the surface projection of two nodal loops at $k_z$ = 0 and $\pi$ without passing through the $\Gamma$ point [Fig. 4(h)], distinct from the drumhead SSs. We discuss in the following that such a difference could be explained by the Zak phase.

It has been theoretically proposed that the topological index of a nodal loop under negligible SOC is given by the Zak phase  which corresponds to a line integral of Berry connection along a path parallel to the $k_z$ axis in the 3D BZ [\onlinecite{ChanPRB2016}], 
\begin{eqnarray}
P(\bm{k}_\parallel) = -i\sum_{E_i<E_{\rm F}} \int_{-\pi}^{\pi} {\langle u_i(\bm{k}) | \partial {k_z} | u_i(\bm{k})\rangle} dk_z
\end{eqnarray}
where $u_i(\bm{k})$ is a Bloch wave function for the $i$th band, as shown by vertical red arrows in Figs. 4(g) and 4(f). The Zak phase takes a different value depending on whether the integral path is inside or outside the nodal loop, which corresponds to the band-inverted or band-non-inverted region, respectively. When a single nodal loop exists [Fig. 4(g)] (e.g. at the $k_z$=0 plane), one can naturally expect an emergence of a drumhead SS at the (001) surface due to the non-trivial Zak phase of $\pi$ inside the loop. When another smaller nodal loop is added to the $k_z$=$\pi$ plane [Fig. 4(h)], the Zak phase inside the small loop becomes 2$\pi$ (= $\pi$ + $\pi$) because the band inversion takes place twice (one for the larger and the other for the smaller loops). Since the Zak phase of 2$\pi$ is equal to 0, we expect the absence of nontrivial SS (nontrivial in terms of the Zak phase but not in terms of the Z$_2$ topological invariant) in this double band-inverted $k$ region. Such absence of the nontrivial SS is also expected for the $k$ region outside the larger nodal loop, because this region is also characterized by the Zak phase of 0 due to the absence of band inversion. On the other hand, in the $k$ region between the smaller and larger nodal loops, the Zak phase becomes $\pi$ due to the single band inversion from the larger nodal loop. As a result, nontrivial SS is expected to emerge. Intriguingly, this is exactly the case of the present ARPES observation which reveals the existence of Dirac-node-arc SS between the $k_z$=0 and $\pi$ nodal loops.

While the above discussion assumes a negligible SOC, an influence of a finite SOC, which opens up an energy gap at the nodal loop, should not be neglected in the actual system, as can be seen from the SOC-gap opening in our band-structure calculations in Figs. 2(b)-2(d). In this context, one may think that the discussion based on the Z$_2$ topological index applied for 3D TI would be more appropriate to specify the topological nature of the Dirac-node arc SS. From a consideration of $k$ location of the band inversion (i.e., $\Gamma$ and {\it Z}), one can conclude that the system becomes a weak TI characterized by the Z$_2$ topological index of (0;001) if we assume a curved chemical potential that passes the $k$-dependent spin-orbit gap. In such a case, a topological SS would not emerge on the (001) surface (if a SS appears, it would be trivial in terms of the Z$_2$ index); however, this is apparently conflicting with our observation. Thus, an argument that simply replies on the Z$_2$ topological index does not catch a whole picture of the underlying topology of the Dirac-node-arc SS. This conclusion is further supported by our observation of universal Dirac-node-arc SS on several different materials (ZrGeS, ZrGeSe, ZrGeTe, and HfSiS) that suggests its robustness against structural perturbations; this cannot be simply explained in terms of the trivial nature of the Dirac-node-arc SS. The present result thus suggests that, even under a finite SOC, the concept of the Zak phase is still useful to discuss the underlying topology of the nodal loops and the Dirac-node-arc SS.
 
\section{SUMMARY}
Our SX- and VUV-ARPES studies of ZrGeSe revealed the existence of bulk nodal loops both at $k_z$ = 0 and $\pi$, together with the Dirac-node-arc SSs connecting the nodal loops. We found that the Dirac-node-arc SS is a universal feature of ZrGeX$_{\rm c}$ (X$_{\rm c}$ = S, Se, and Te) despite the quantitative difference in the shape of nodal loops. This result serves as a foundation for understanding the interplay between the appearance of SSs and bulk nodal lines in LNSMs.

\begin{acknowledgments}
We thank A. Yamakage for fruitful discussions. This work was funded by JST-CREST (No: JPMJCR18T1), JST-PRESTO (No: JPMJPR18L7), MEXT of Japan (Innovative Area ``Topological Materials Science'' JP15H05853 and JP15K21717), JSPS (JSPS KAKENHI No: JP17H01139, JP26287071, JP25220708, JP18H01160, JP18H04472, and JP18J20058), KEK-PF (Proposal number: 2018S2-001, 2016G-555, and 2018G-653), and by the Deutsche Forschungsgemeinschaft (DFG, German Research Foundation) - Project No. 277146847 - CRC 1238 (Subproject A04).
\end{acknowledgments}

\appendix
\section{SPIN-ORBIT-COUPLING EFFECT ON BULK NODAL LINE IN ZrGeSe}
We discuss in Fig. 5 the influence of SOC on the electronic band structure of ZrGeSe. ZrGeSe is expected to show a stronger SOC effect compared to the ZrSiX$_{\rm c}$ family [\onlinecite{SchoopNC2016,ChenPRB2017,NeupanePRB2016,ToppNJP2016,HosenPRB2017,floatingPRX2017}] due to heavier atomic mass of Ge than that of Si. The inclusion of SOC in the bulk-band calculation of ZrGeSe resulted in the energy gap opening of maximally 30 meV at the nodal point, as shown in Figs. 5(a) and 5(b). Similar behavior is also seen in the slab calculations for 4QL ZrGeSe [Figs. 5(c) and 5(d)].

\begin{figure}
\includegraphics[width=3.2in]{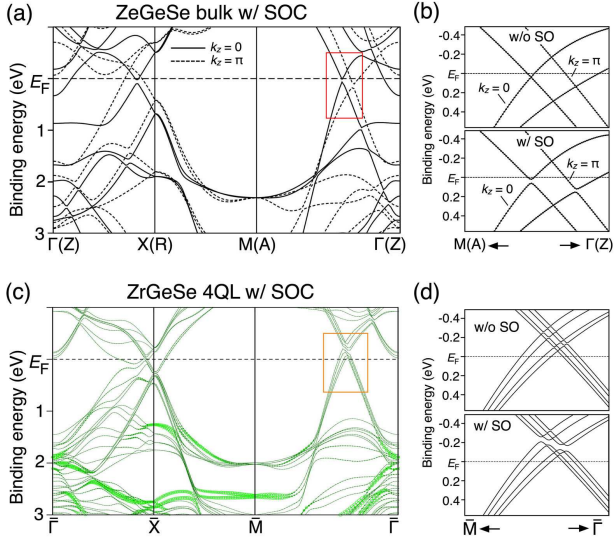}
\hspace{0.2in}
\vspace{-0.5 cm}
\caption{(color online). (a) Calculated band structure including SOC for bulk ZrGeSe. (b) Magnified view of the band structure in the area enclosed by red rectangle in (a). Upper and lower panel shows the case without and with SOC. (c), (d) Same as (a) and (b) but for 4QL slab.
}
\vspace{-0.5 cm}
\end{figure}

We show in Figs. 6(a) and 6(b) the ARPES-intensity mapping as a function of 2D wave vector ($k_x$ and $k_y$) for some representative binding-energy slices ($E_{\rm B}$ = 0, 0.4, and 0.8 eV), measured at $k_z$ = 0 and $\pi$ planes, respectively, with SX photons. We also show in Figs. 6(c) and 6(d) the calculated bulk-band dispersion with and without SOC, respectively, obtained along several $k$ cuts in the $k_x$-$k_y$ plane across the nodal loops [cuts A-F in Fig.6 (a) and 6(b)]. As shown by the calculated band dispersion without SOC along cuts A-C in Fig. 6(c), the nodal points (band-crossing points) exist very close to $E_{\rm F}$ at $k_z$ = 0, whereas those at $k_z$ = $\pi$ are located 0.15--0.2 eV below $E_{\rm F}$. Upon inclusion of SOC, the nodal points are gapped with maximally 30 meV as shown in Fig. 6(d), while the overall linear dispersion outside the gap remains almost unchanged. As seen in Fig. 6(a), a diamond-like intensity pattern at $E_{\rm B}$ = 0 eV which originates from nearly degenerate Ge 4{\it p} and Zr 4{\it d} bands, splits into inner (Zr 4{\it d}) and outer (Ge 4{\it p}) diamonds upon increasing $E_{\rm B}$, and they gradually separate from each other, consistent with the overall calculated band dispersion in Figs. 6(c) and 6(d). A similar trend is also seen for the band dispersion at $k_z$ = $\pi$ in Fig. 6(b), whereas a small splitting of diamonds at $E_{\rm B}$ = $E_{\rm F}$ expected from the calculation is not clearly resolved likely due to insufficient resolution of SX-ARPES.

\begin{figure}
\includegraphics[width=3.45 in]{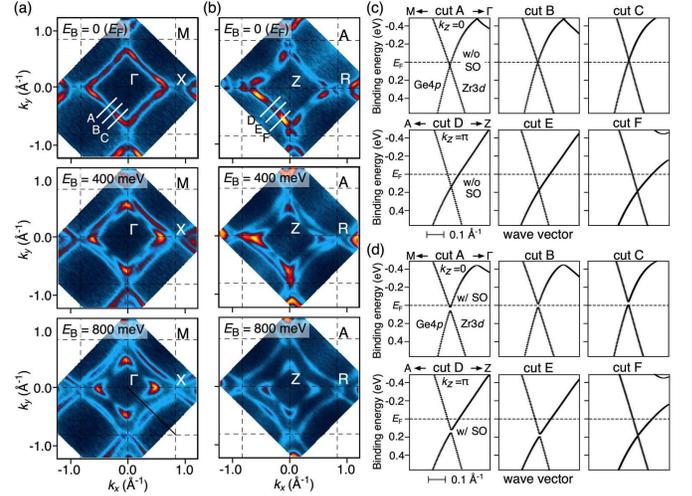}
\hspace{0in}
\caption{(color online). (a), (b) ARPES-intensity mapping of ZrGeSe at $k_z$ = 0 and $\pi$ respectively, plotted as a function of $k_x$ and $k_y$ for representative $E_{\rm B}$ slices ($E_{\rm B}$ = 0, 0.4, and 0.8 eV), measured with SX photons. (c), (d) Calculated band structure without and with SOC, respectively, obtained along cuts A-F in (a) and (b).
}
\end{figure}

\begin{figure*}
\includegraphics[width=6.4 in]{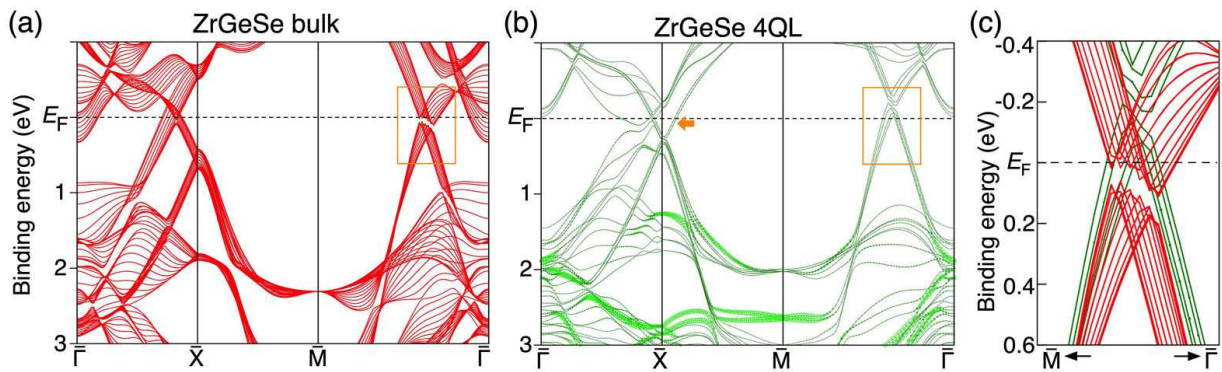}
\hspace{0.2in}
\caption{(color online).  Calculated band dispersion of ZrGeSe for (a) bulk and (b) 4QL slab. The 4QL slabs are separated by 1.5-nm-thick vacuum layer. Thickness of green lines in (b) represents a weight of atomic orbitals projected onto topmost QL. (c) Direct comparison of band dispersion around the nodal line in the region enclosed by orange rectangle in (a) and (b). During the calculation, atomic positions of the slab are relaxed while keeping crystal symmetry. Also, in-plane lattice constant is fixed to the experimental value ($a$ = 3.706 \AA). 
}
\end{figure*}

\section{COMPARISON OF BULK AND SLAB CALCULATIONS FOR ZrGeSe}
To discuss the surface contribution to the electronic states, we show in Figs. 7(a) and 7(b) the calculated band structure for ZrGeSe along the high-symmetry lines for bulk at several $k_z$ slices, and for 4 QL slab, respectively. As can be seen from the area enclosed by an orange rectangle, the calculations for bulk and 4QL commonly show narrow spin-orbit gap around the crossing points (line node) of the linearly dispersive bands. Interestingly, the gap in the slab calculation is located at a higher energy position by $\sim$0.25 eV compared to that in the bulk-band calculation. Such a difference likely originates from a change in the charge balance associated with an emergence of trivial SS near $E_{\rm F}$ around the $\bar{X}$ point in the slab calculation (marked by orange arrow). When we directly overlap these calculations in Fig. 7(c), the bands forming the lower Dirac-like dispersion in the slab calculation appear to accidentally disperse across the bulk band gap due to the mismatch in the location of the energy gap. We have examined our slab calculations and found that the ``SS" traversing the bulk-band gap in Fig. 7(c) show ordinary dispersion in the $k_x$-$k_y$ plane, different from the dispersion expected from the Dirac-node-arc-type dispersion.

\begin{figure}[b]
\includegraphics[width=3in]{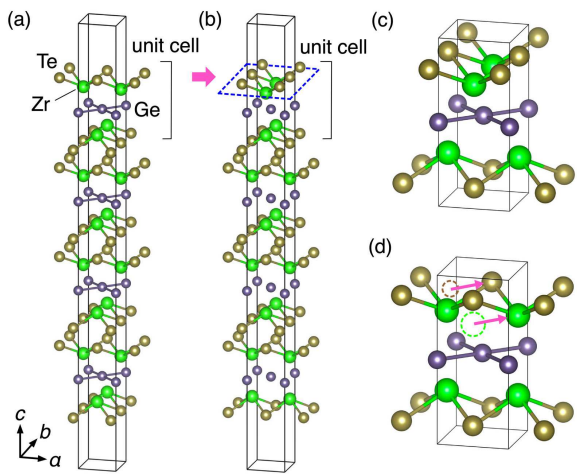}
\hspace{0.2in}
\caption{(color online).  (a) 4QL ZrGeTe slab with 1.5-nm vacuum layer. (b) Modified slab with the SF at the top and bottom Zr-Te layers. (c), (d) Magnified views of the topmost unit cell for (a) and (b), respectively. 
}
\end{figure}

\section{SLAB CALCULATIONS FOR ZrGeTe WITH VARIOUS STRUCTURE MODELS}
To theoretically reproduce the Dirac-node-arc SS, we have considered 4QL-thick ZrGeTe with 1.5-nm vacuum layer [see Fig. 8(a)] as a basic structure, and examined various possibilities by modifying that structure. We have chosen ZrGeTe because the SOC is the strongest among ZrGeX$_{\rm c}$ so that the change in the electronic states can be most clearly visible.

We first examined a stacking fault (SF) at the surface. We assumed a slab [Fig. 8(b)] modified from the original slab [Fig. 8(a)] that contains the top and bottom Zr--Te layers shifted by (1/2 {\bf G}, 1/2 {\bf G}, 0) ({\bf G}: reciprocal lattice vector) from the original slab. This shift is directly visible from a comparison of the topmost unit layer without SF [Fig. 8(c)] and with SF [Fig. 8(d)]. We show in Figs. 8(a) and 8(b) the corresponding calculated band structures. One can immediately see that the X-shaped Dirac-like band at $E_{\rm B} \sim$ 0.3 eV in the SF-free calculation, which is attributed to the trivial SS [\onlinecite{SchoopNC2016,ChenPRB2017,NeupanePRB2016,TakanePRB2016,ToppNJP2016,HosenPRB2017,floatingPRX2017,HosenPRB2018}], is not visible in the calculation with the SF. Instead, a Dirac-like band appears at slightly right-hand side of the $\bar{X}$  point (midway between $\bar{X}$ and $\bar{M}$). Since this feature is absent in the experiment, one can say at this stage that such SF does not occur. A comparison of the band dispersions along the $\overline{\Gamma}\overline{M}$ cut between Figs. 9(a) and 9(b) also reveals that the overall gap magnitude is smaller in the calculation with SF compared to that without SF. This is more clearly visible from a magnified view around the nodal line [area enclosed by red rectangle in Figs. 9(a) and 9(b)] in Figs. 9(c) and 9(d). While the reduced energy gap in the calculation with SF may be in favor of the gapless nature of the SS seen in the experiment, we found that the calculated band structure hardly reproduces the overall Dirac-node-arc dispersion, in particular, a sudden change in the band velocity. These results suggest that the SF is unlikely to be the origin of the Dirac-node-arc SS.

Next we considered the possibility of surface relaxations. Overall, the bulk and slab calculations presented in this study were carried out by relaxing atomic positions (e.g., calculations in Fig. 7 are for the relaxed lattice). To examine the effect of surface relaxations, we have carried out the slab calculation by fixing the atomic positions and lattice constants as those in the experimental bulk crystal structure, and the result is shown in Fig. 9(e). When this is compared with the normal calculation with surface relaxations [Fig. 9(a)], one can see no obvious difference between the two, in particular, regarding the overall band dispersion around the nodal line along the $\overline{\Gamma}\overline{M}$ cut. This suggests that the surface relaxations do not play a crucial role in the emergence of the Dirac-node-arc SS.

\begin{figure*}
\includegraphics[width=6.4in]{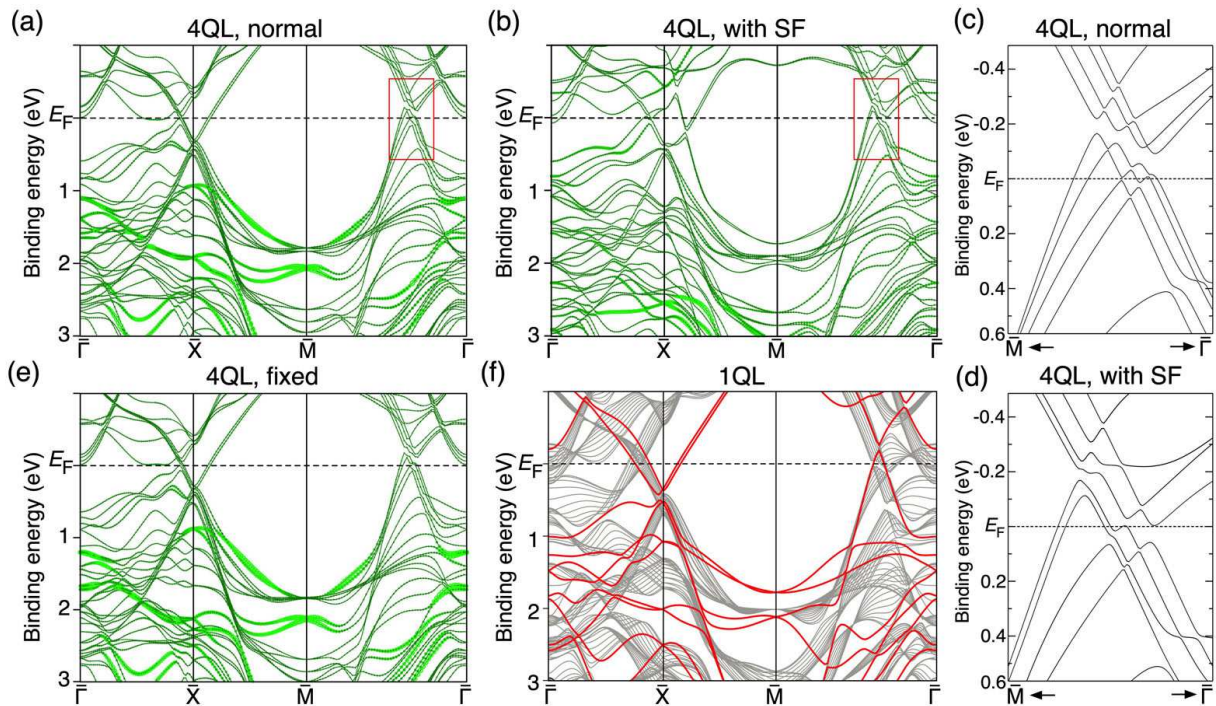}
\hspace{0.2in}
\caption{(color online). (a) Calculated band structure for 4QL ZrGeTe slab. Thickness of green lines represents a weight of atomic orbitals projected onto topmost QL. (b) Same as (a) but for modified slab with SF [see Fig. R3(b) for the slab structure]. (c), (d) Expanded view of band structure in the area enclosed by red rectangle in (a) and (b), respectively. (e) Same as (a) but calculated with fixed input parameters of atomic positions. (f) Calculated band structure for a free-standing 1QL slab of ZrGeTe (red curves), overlaid with the projection of calculated bulk bands (gray curves).
}
\end{figure*}

We also considered the possibility that the topmost layer is isolated from the bulk and forms a 1QL-like energy bands, by referring to the previous study that reported a similarity between the experimental surface band and the calculated 1QL band [\onlinecite{LouPRB2016}]. As shown in Fig. 9(f), our calculation for the 1QL slab shows a Dirac-like band along the $\overline{\Gamma}\overline{M}$ cut. However, this band hardly reproduces the observed sizable change in the band velocity with respect to that of the bulk, a characteristic of the Dirac-node-arc SS.

All these considerations suggest that the observed Dirac-node-arc SS is hardly reproduced by the slab calculations, even when some structural modifications are taken into account. We speculate that the reason why the slab calculation hardly reproduces the observed Dirac-node-arc SS is that the bulk spin-orbit gap is very small. This enforces the SS to be located around the edge of the bulk band, resulting in a long penetration depth into the bulk. Unlike the case of a real crystal, a finite slab size in the calculation would result in band hybridization between the top and bottom SS and may strongly modulate the shape of SS. However, such hybridization should not occur in the real crystal with $\mu$m-size thickness. In such a case, the band structure obtained with the slab calculation would hardly reproduce the experimental band structure. We therefore think that an inherent difficulty in properly treating the SS inside the small bulk spin-orbit gap prohibits the slab calculations from correctly predicting the Dirac-node-arc SS.

\bibliographystyle{prsty}

\begin{thebibliography}{60}

\bibitem{HasanRMP2010} M. Z. Hasan and C. L. Kane, Rev. Mod. Phys. \textbf{82}, 3045 (2010).

\bibitem{ZhangRMP2011} X.-L. Qi and S.-C. Zhang, Rev. Mod. Phys. \textbf{83}, 1057 (2011).

\bibitem{AndoJPSJ2013} Y. Ando, J. Phys. Soc. Jpn. \textbf{82}, 102001 (2013).

\bibitem{FuPRL2011} L. Fu, Phys. Rev. Lett. \textbf{106}, 106802 (2011). 

\bibitem{HsiehNC2012} T. H. Hsieh, H. Lin, J. Liu, W. Duan, A. Bansil, and L. Fu, Nat. Commun. \textbf{3}, 982 (2012). 

\bibitem{TanakaNP2012} Y. Tanaka, Z. Ren, T. Sato, K. Nakayama, S. Souma, T. Takahashi, K. Segawa, and Y. Ando, Nat. Phys. \textbf{8}, 800 (2012). 

\bibitem{XuNC2012} S.-Y. Xu, C. Liu, N. Alidoust, M. Neupane, D. Qian, I. Belopolski, J. D. Denlinger, Y. J. Wang, H. Lin, L. A. Wray, G. Landolt, B. Slomski, J. H. Dill, A. Marcinkova, E. Morosan, Q. Gibson, R. Sankar, F. C. Chou, R. J. Cava, A. Bansil, and M. Z. Hasan, Nat. Commun. \textbf{3}, 1192 (2012). 

\bibitem{DziawaNM2012} P. Dziawa, B. J. Kowalski, K. Dybko, R. Buczko, A. Szczerbakow, M. Szot, E. {\L}usakowska, T. Balasubramanian, B. M. Wojek, M. H. Berntsen, O. Tjernberg, and T. Story, Nat. Mater. \textbf{11}, 1023 (2012). 

\bibitem{LFuPRL2010} L. Fu and E. Berg, Phys. Rev. Lett. \textbf{105}, 097001 (2010).

\bibitem{SasakiPRL2011} S. Sasaki, M. Kriener, K. Segawa, K. Yada, Y. Tanaka, M. Sato, and Y. Ando, Phys. Rev. Lett. \textbf{107}, 217001 (2011).

\bibitem{NMRNP2016} K. Matano, M. Kriener, K. Segawa, Y. Ando, and G.-q. Zheng, Nat. Phys. \textbf{12}, 852 (2016). 

\bibitem{SasakiPRL2012} S. Sasaki, Z. Ren, A. A. Taskin, K. Segawa, L. Fu, and Y. Ando, Phys. Rev. Lett. \textbf{109}, 217004 (2012).

\bibitem{Sato2017} M. Sato and Y. Ando, Rep. Prog. Phys. \textbf{80}, 076501 (2017).

\bibitem{WengCM2016} H. Weng, X. Dai, and Z. Fang, J. Phys.: Condens. Matter \textbf{28}, 303001 (2016).

\bibitem{FangCPB2016} C. Fang, H. Weng, X. Dai, and Z. Fang, Chin. Phys. B \textbf{25}, 117106 (2016).

\bibitem{WangAPX2017} S. Wang, B.-C. Lin, A.-Q. Wang, D. P. Yu, and Z. M. Liao, Adv. Phys.: X \textbf{2}, 518 (2017).

\bibitem{ArmitageRMP2018} N. P. Armitage, E. J. Mele, and A. Vishwanath, Rev. Mod. Phys. \textbf{90}, 015001 (2018). 

\bibitem{BernevigJPSJ2018} A. Bernevig, H. Weng, Z. Fang, and X. Dai, J. Phys. Soc. Jpn. \textbf{87}, 041001 (2018).

\bibitem{MurakamiJPSJ2018} M. Hirayama, R. Okugawa, and S. Murakami, J. Phys. Soc. Jpn. \textbf{87}, 041002 (2018).

\bibitem{SchoopCM2018} L. M. Schoop, F. Pielnhofer, and B. V. Lotsch, Chem. Mater. \textbf{30}, 3155 (2018).

\bibitem{YangARX2018} S.-Y. Yang, H. Yang, E. Derunova, S. S. P. Parkin, B. Yan, and M. N. Ali, Adv. Phys.: X \textbf{3}, 1414631 (2018).

\bibitem{WanPRB2011} X. Wan, A. M. Turner, A. Vishwanath, and S. Y. Savrasov, Phys. Rev. B \textbf{83}, 205101 (2011).

\bibitem{YangPRB2011} K.-Y. Yang, Y.-M. Lu, and Y. Ran, Phys. Rev. B \textbf{84}, 075129 (2011).

\bibitem{HaldaneArXiv2014} F. D. M. Haldane, arXiv:1401.0529 (2014).

\bibitem{MurakamiPRB2014} R. Okugawa and S. Murakami, Phys. Rev. B \textbf{89}, 235315 (2014).

\bibitem{XuSci2015} S.-Y. Xu, I. Belopolski, N. Alidoust, M. Neupane, G. Bian, C. Zhang, R. Sankar, G. Chang, Z. Yuan, C.-C. Lee, S.-M. Huang, H. Zheng, J. Ma, D. S. Sanchez, B. K. Wang, A. Bansil, F. Chou, P. P. Shibayev, H. Lin, S. Jia, and M. Z. Hasan, Science \textbf{349}, 613 (2015).

\bibitem{BQLuPRX2015} B. Q. Lv, H. M. Weng, B. B. Fu, X. P. Wang, H. Miao, J. Ma, P. Richard, X. C. Huang, L. X. Zhao, G. F. Chen, Z. Fang, X. Dai, T. Qian, and H. Ding, Phys. Rev. X \textbf{5}, 031013 (2015).

\bibitem{YLChenNP2015} L. X. Yang, Z. K. Liu, Y. Sun, H. Peng, H. F. Yang, T. Zhang, B. Zhou, Y. Zhang, Y. F. Guo, M. Rahn, D. Prabhakaran, Z. Hussain, S.-K. Mo, C. Felser, B. Yan, and Y. L. Chen, Nat. Phys. \textbf{11}, 728 (2015).

\bibitem{BQLuPRL2015} B. Q. Lv, S. Muff, T. Qian, Z. D. Song, S. M. Nie, N. Xu, P. Richard, C. E. Matt, N. C. Plumb, L. X. Zhao, G. F. Chen, Z. Fang, X. Dai, J. H. Dil, J. Mesot, M. Shi, H. M. Weng, and H. Ding, Phys. Rev. Lett. \textbf{115}, 217601 (2015).

\bibitem{XuPRL2016} S.-Y. Xu, I. Belopolski, D. S. Sanchez, M. Neupane, G. Chang, K. Yaji, Z. Yuan, C. Zhang, K. Kuroda, G. Bian, C. Guo, H. Lu, T.-R. Chang, N. Alidoust, H. Zheng, C.-C. Lee, S.-M. Huang, C.-H. Hsu, H.-T. Jeng, A. Bansil, T. Neupert, F. Komori, T. Kondo, S. Shin, H. Lin, S. Jia, and M. Z. Hasan, Phys. Rev. Lett. \textbf{116}, 096801 (2016).

\bibitem{SoumaPRB2016} S. Souma, Z. Wang, H. Kotaka, T. Sato, K. Nakayama, Y. Tanaka, H. Kimizuka, T. Takahashi, K. Yamauchi, T. Oguchi, K. Segawa, and Y. Ando, Phys. Rev. B \textbf{93}, 161112(R) (2016).

\bibitem{SchoopNC2016} L. M. Schoop, M. N. Ali, A. Topp, A. Varykhalov, D. Marchenko, V. Duppel, S. S. P. Parkin, B. V. Lostch, and C. R. Ast, Nat. Commun. \textbf{7}, 11696 (2016).

\bibitem{ChenPRB2017} C. Chen, X. Xu, J. Jiang, S.-C. Wu, Y. P. Qi, L. X. Yang, M. X. Wang, Y. Sun, N. B. M. Schr\"{o}ter, H. F. Yang, L. M. Schoop, Y. Y. Lv, J. Zhou, Y. B. Chen, S. H. Yao, M. H. Lu, Y. F. Chen, C. Felser, B. H. Yan, Z. K. Liu, and Y. L. Chen, Phys. Rev. B \textbf{95}, 125126 (2017).

\bibitem{NeupanePRB2016} M. Neupane, I. Belopolski, M. M. Hosen, D. S. Sanchez, R. Sankar, M. Szlawska, S.-Y. Xu, K. Dimitri, N. Dhakal, P. Maldonado, P. M. Oppeneer, D. Kaczorowski, F. Chou, M. Z. Hasan, and T. Durakiewicz, Phys. Rev. B \textbf{93}, 201104(R) (2016). 

\bibitem{LouPRB2016} R. Lou, J.-Z. Ma, Q.-N. Xu, B.-B. Fu, L.-Y. Kong, Y.-G. Shi, P. Richard, H.-M. Weng, Z. Fang, S.-S. Sun, Q. Wang, H.-C. Lei, T. Qian, H. Ding, and S.-C. Wang, Phys. Rev. B \textbf{93}, 241104(R) (2016).

\bibitem{TakanePRB2016} D. Takane, Z. Wang, S. Souma, K. Nakayama, C. X. Trang, T. Sato, T. Takahashi, and Y. Ando, Phys. Rev. B \textbf{94}, 121108(R) (2016).

\bibitem{ToppNJP2016} A. Topp, J. M. Lippmann, A. Varykhalov, V. Duppel, B. V. Lostch, C. R. Ast, and L. M. Schoop, New J. Phys. \textbf{18}, 125014 (2016).


\bibitem{HosenPRB2017} M. M. Hosen, K. Dimitri, I. Belopolski, P. Maldonado, R. Sankar, N. Dhakal, G. Dhakal, T. Cole, P. M. Oppeneer, D. Kaczorowski, F. Chou, M. Z. Hasan, T. Durakiewicz, and M. Neupane, Phys. Rev. B \textbf{95}, 161101(R) (2017).

\bibitem{floatingPRX2017} A. Topp, R. Queiroz, A. Gr\"uneis, L. M\"uchler, A. W. Rost, A. Varykhalov, D. Marchenko, M. Krivenkov, F. Rodolakis, J. L. McChesney, B. V. Lotsch, L. M. Schoop, and C. R. Ast, Phys. Rev. X \textbf{7}, 041073 (2017).

\bibitem{HosenPRB2018} M. M. Hosen, K. Dimitri, A. Aperis, P. Maldonado, I. Belopolski, G. Dhakal, F. Kabir, C. Sims, M. Z. Hasan, D. Kaczorowski, T. Durakiewicz, P. M. Oppeneer, and M. Neupane, Phys. Rev. B \textbf{97}, 121103(R) (2018).

\bibitem{ZrSiSCalc2015} Q. Xu, Z. Song, S. Nie, H. Weng, Z. Fang, and X. Dai, Phys. Rev. B \textbf{92}, 205310 (2015).

\bibitem{Kumar2017}  N. Kumar, K. Manna, Y. Qi, S.-C. Wu, L. Wang, B. Yan, C. Felser, and C. Shekhar, Phys. Rev. B \textbf{95}, 121109 (2017).

\bibitem{VASP} G. Kresse and J. Furthm\"{u}ller,  Phys. Rev. B \textbf{54}, 11169 (1996).

\bibitem{GGA}  J. P. Perdew, K. Burke and M. Ernzerhof, Phys. Rev. Lett.  \textbf{77}, 3865 (1996).

\bibitem{FuSXarXiv2017} B.-B. Fu, C.-J. Yi, T.-T. Zhang, M. Caputo, J.-Z. Ma, X. Gao, B. Q. Lv, L.-Y. Kong, Y.-B. Huang, P. Richard, M. Shi, V. N. Strocov, C. Fang, H.-M. Weng, Y.-G. Shi, T. Qian, and H. Ding, Sci. Adv. \textbf{5}, eaau6459 (2019).

\bibitem{SuNJP2018} C. C. Su, C. S. Li, T. C. Wang, S. Y. Guan, R. Sankar, F. Chou, C. S. Chang, W. L. Lee, G. Y. Guo, and T. M. Chuang, New J. Phys. \textbf{20}, 103025 (2018).

\bibitem{LodgeNL2017} M. S. Lodge, G. Chang, C. Y. Huang, B. Singh, J. Hellerstedt, M. T. Edmonds, D. Kaczorowski, M. M. Hosen, M. Neupane, H. Lin, M. S. Fuhrer, B. Weber, and M. Ishigami, Nano Lett. \textbf{17}, 7213 (2017).

\bibitem{ZrSiSNP2017} S. Pezzini, M. R. van Delft, L. M. Schoop, B. V. Lotsch, A. Carrington, M. I. Katsnelson, N. E. Hussey, and S. Wiedmann, Nat. Phys. \textbf{14}, 178 (2017).

\bibitem{RudenkoPRL2018} A. N. Rudenko, E. A. Stepanov, A. I. Lichtenstein, and M. I. Katsnelson, Phys. Rev. Lett. \textbf{120}, 216401 (2018).

\bibitem{HirayamaNC2017} M. Hirayama, R. Okugawa, T. Miyake, and S. Murakami, Nat. Commun. \textbf{8}, 14022 (2017).

\bibitem{AhmPRL2018} J. Ahn, D. Kim, Y. Kim, and B.-J. Yang, Phys. Rev. Lett. \textbf{121}, 106403 (2018).

\bibitem{FangPRB2015} C. Fang, Y. Chen, H.-Y. Kee, and L. Fu, Phys. Rev. B \textbf{92}, 081201(R) (2015).

\bibitem{YamakageJPSJ2016} A. Yamakage, Y. Yamakawa, Y. Tanaka, and Y. Okamoto, J. Phys. Soc. Jpn. \textbf{85}, 013708 (2016).

\bibitem{ChanPRB2016} Y.-H. Chan, C.-K. Chiu, M. Y. Chou, and A. P. Schnyder, Phys. Rev. B \textbf{93}, 205132 (2016).

\bibitem{BurkovPRB2011} A. A. Burkov, M. D. Hook, and L. Balents, Phys. Rev. B \textbf{84}, 235126 (2011)

\bibitem{KimPRL2015} Y. Kim, B. J. Wieder, C. L. Kane, and A. M. Rappe, Phys. Rev. Lett. \textbf{115}, 036806 (2015).

\bibitem{BianPRB2016} G. Bian, T.-R. Chang, H. Zheng, S. Velury, S.-Y. Xu, T. Neupert, C.-K. Chiu, S.-M. Huang, D. S. Sanchez, I. Belopolski, N. Alidoust, P.-J. Chen, G. Chang, A. Bansil, H.-T. Jeng, H. Lin, and M. Z. Hasan, Phys. Rev. B \textbf{93}, 121113(R) (2016).
 
\end{thebibliography}

\end{document}